\documentclass[fleqn,10pt]{wlscirep}
\usepackage[utf8]{inputenc}
\usepackage[T1]{fontenc}

\usepackage{fancyref}
\usepackage{pdfpages}
\usepackage{mathptmx}

\usepackage{algorithm}
\usepackage{algorithmic}
\usepackage{color}
\usepackage{xcolor}
\usepackage{caption}
\usepackage{hyperref}

\usepackage{subcaption}
\usepackage{array}
\usepackage{tabularray}
\usepackage{cite} 

\usepackage{graphicx} 
\usepackage{amsmath} % From the American Mathematical Society
\usepackage{multirow}
\usepackage{acro}

\DeclareAcronym{op}{
  short = OP ,
  long  = Osteoporosis ,
  sort  = OP ,
}
\DeclareAcronym{oa}{
  short = OA ,
  long  = Osteoarthritis ,
  sort  = OA ,
}
\DeclareAcronym{msk}{
  short = MSK ,
  long  = Musculoskeletal ,
  sort  = MSK ,
}
\DeclareAcronym{cnn}{
  short = CNN ,
  long  = Convolutional Neural Network ,
  sort  = CNN ,
}
\DeclareAcronym{ao}{
  short = AO ,
  long  = Adaptive Optics ,
  sort  = AO ,
}
\DeclareAcronym{dnn}{
  short = DNN ,
  long  = Deep Neural Network ,
  sort  = DNN ,
}
\DeclareAcronym{psf}{
  short = PSF ,
  long  = Point Spread Function ,
  sort  = PSF ,
}
\DeclareAcronym{mse}{
  short = MSE ,
  long  = Mean Squared Error ,
  sort  = MSE ,
}
\DeclareAcronym{rmse}{
  short = RMSE ,
  long  = Root Mean Squared Error ,
  sort  = RMSE ,
}

\DeclareAcronym{ssim}{
  short = SSIM ,
  long  = Structural Similarity Index Measure ,
  sort  = SSIM ,
}

\DeclareAcronym{psnr}{
  short = PSNR ,
  long  = Peak Signal-to-Noise Ratio ,
  sort  = PSNR ,
}

\title{Direct Zernike Coefficient Prediction from Point Spread Functions and Extended Images using Deep Learning}

\author[1,*]{Yong En Kok}
\author[2]{Alexander Bentley}
\author[1]{Andrew Parkes}
\author[2]{Amanda J. Wright}
\author[2,3]{Michael G. Somekh}
\author[1]{Michael Pound}
\affil[1]{School of Computer Science, University of Nottingham, Nottingham, NG8 1BB, UK}
\affil[2]{Optics and Photonics Group, Department of Electrical and Electronic Engineering, University of
Nottingham, Nottingham, NG7 2RD, UK}
\affil[3]{Research Center for Humanoid Sensing, Zhejiang Laboratory Hangzhou 3111100, China}

\affil[*]{yong.kok@nottingham.ac.uk}

\begin{abstract}
Optical imaging quality can be severely degraded by system and sample induced aberrations. Existing adaptive optics systems typically rely on iterative search algorithm to correct for aberrations and improve images. This study demonstrates the application of convolutional neural networks to characterise the optical aberration by directly predicting the Zernike coefficients from two to three phase-diverse optical images. We evaluated our network on 600,000 simulated \ac{psf} datasets randomly generated within the range of -1 to 1 radians using the first 25 Zernike coefficients. The results show that using only three phase-diverse images captured above, below and at the focal plane with an amplitude of 1 achieves a low RMSE of 0.10 radians on the simulated \ac{psf} dataset. Furthermore, this approach directly predicts Zernike modes simulated extended 2D samples, while maintaining a comparable RMSE of 0.15 radians. We demonstrate that this approach is effective using only a single prediction step, or can be iterated a small number of times. This simple and straightforward technique provides rapid and accurate method for predicting the aberration correction using three or less phase-diverse images, paving the way for evaluation on real-world dataset.

\end{abstract}

\begin{document}  

\flushbottom
\maketitle

\thispagestyle{empty}


\section{Introduction}
Optical imaging quality can be adversely affected by wavefront aberrations arising from both system-induced and sample-induced factors. System induced aberrations, stemming from imperfections in system manufacturing or misalignment during assembly, can degrade overall picture quality. Sample induced aberrations are specific to the sample being imaged. Variations in absorption and complex refractive index distributions within the sample, as described by Schwertner et al. \cite{schwertner2004measurement}, lead to reduced signal intensity and image quality. Furthermore, as we image deep into tissue, the presence of multiple scattering processes introduces more intricate aberrations with increasing high-order components. These aberrations will further distort the wavefront, leading to a degraded imaging focus and consequently impose limitations on deep tissue imaging \cite{booth2014adaptive}.

To address optical aberration, \ac{ao} has been introduced to cancel out the aberrations present within the system \cite{girkin2009adaptive,booth2014adaptive}. In an \ac{ao} system, aberrations are corrected using a wavefront correction device to shape the wavefront of the light (eg. deformable mirror \cite{zhu1999adaptive} or a spatial light modulator). To obtain the appropriate wavefront required for aberration correction, there are two main methods: wavefront sensing and optimisation. The former measures the aberration using a wavefront sensor (e.g. a Shack Hartmann sensor \cite{platt2001history}, an interferometer \cite{hardy1998adaptive} or a modal wavefront sensor \cite{neil2000new}). This approach offers high speed and accuracy, but requires a guide-star which can perturb the biological sample. By comparison, the optimization method searches for the appropriate wavefront correction by using a search algorithm such as stochastic parallel gradient descent algorithm (SPGD) \cite{vorontsov2000adaptive}, simulated annealing algorithm \cite{zommer2006simulated}, genetic algorithm \cite{yang2007intracavity}, Gerchberg-Saxton \cite{gerchberg1972practical} or the Extended Nijboer-Zernike theorem \cite{janssen2002extended}. These conventional search algorithms can be iterative, slow and computationally demanding. Speed is a particular problem when working with biological samples where it is important to limit photon dose to reduce the risk of photo-damage, moreover samples are often not static so it is important to make the correction before the aberrations change substantially. 

With the advancement of computer vision, deep learning has emerged as a promising technique for addressing a range of image-based inverse problems by learning the corresponding mapping relationships. A comprehensive review summarizing the advances of classic and modern computational \ac{ao} techniques applied to optical microscopy can be found in \cite{zhang2023adaptive}. Researchers have applied deep learning approaches to perform aberration retrieval by either directly reconstructing the phase of the aberrated wavefront \cite{liu2019deep,guo2019improved} or by recovering the Zernike coefficients \cite{nishizaki2019deep,xu2019improved,ma2019numerical,andersen2019neural,andersen2020image,wu2020sub,krishnan2020optical,saha2020practical,ma2023turbulence,shohani2023using,xin2019object,hu2023universal} from the input aberrated images, followed by the calculation of the aberration phase represented by Zernike modes. By characterising the aberration present, it becomes possible to correct the distorted image using a system such as \ac{ao}, that introduces an equal yet opposite aberration. Nevertheless, deep learning models can encounter challenges in aberration retrieval due to the inherent ambiguity associated with determining the Zernike coefficients solely from an aberrated image without phase information. In particular, angularly-even Zernike polynomials can produce identical \ac{psf} images for oppositely-signed Zernike coefficients, thus posing difficulties for the deep learning models due to the ambiguous relationship \cite{siddik2023deep,saha2020practical}. To address this issue, Siddik \textit{et al}.\cite{siddik2023deep} proposed to predict the absolute values for the specific Zernike polynomials that are susceptible to this ambiguity. However, this method does not represent a complete wavefront sensing approach, because the sign of the coefficient is not considered. Alternatively, some studies \cite{nishizaki2019deep,xu2019improved} have explored the use of a single image acquired through preconditioners such as overexposed, defocused, or scattered images to enhance estimation accuracy.  Nonetheless, these methods \cite{nishizaki2019deep,xu2019improved} still have large aberration measurement errors. To overcome these limitations, researchers \cite{ma2019numerical,guo2019improved,andersen2019neural,andersen2020image,wu2020sub,krishnan2020optical,saha2020practical,ma2023turbulence,shohani2023using} have leveraged phase-diverse images obtained from focus and defocus planes (using $\ge2$ images) to further improve prediction accuracy. While effective for point-like objects, these approaches often struggle to generalize well to extended 2D samples. In contrast, other methods  \cite{xin2019object,hu2023universal} have incorporated physical principles and imaging processes to separate \ac{psf}-related information from the pairs of images, thus enabling phase retrieval from non point-like objects. 

So far, previous studies have primarily focused on simulated datasets with a small number of Zernike coefficients considered and exhibited limitations in prediction accuracy. Our work aims to directly predict 25 Zernike coefficients representing wavefront aberrations applied to simulated data of \ac{psf} and extended 2D sample dataset with high accuracy. We begin by examining the different bias modes, amplitude of bias mode, and the minimum number of phase-diverse images required for a satisfactory prediction on the aberrated \ac{psf}. With these optimal parameters obtained from the evaluation on the aberrated \ac{psf} dataset, we  therefore explore the ability of the network to recover aberrations from extended 2D sources.

\section{Materials and Methodology}

\subsection{Dataset} 
\begin{figure}[ht]
    \centering
    \begin{subfigure}[ht]{0.32\textwidth}
        \centering
        \includegraphics[width=\textwidth]{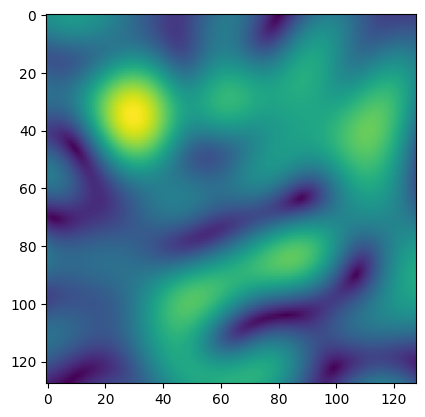}
        \caption{Negative defocus, $I_{-1}$}
        \label{fig:psfSNegativeDefocus}
    \end{subfigure}
    \hfill
    \begin{subfigure}[ht]{0.32\textwidth}
        \centering
        \includegraphics[width=\textwidth]{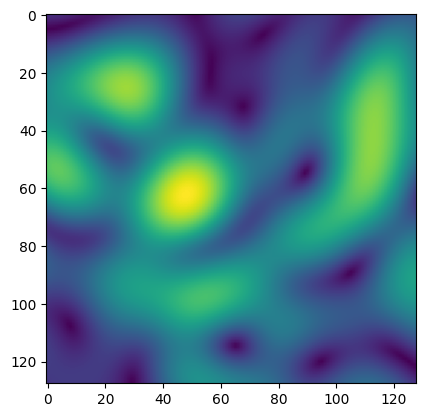}
        \caption{At focus, $I_{0}$}
        \label{fig:psfSAtFocus}
    \end{subfigure}
    \begin{subfigure}[ht]{0.32\textwidth}
        \centering
        \includegraphics[width=\textwidth]{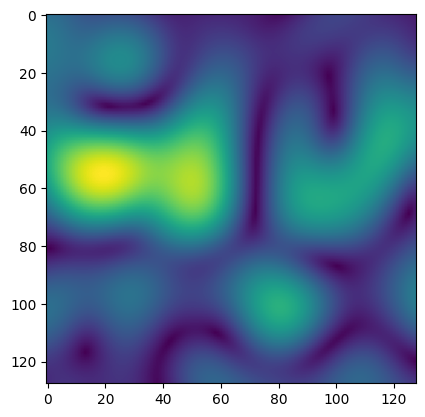}
        \caption{Positive defocus, $I_{+1}$}
        \label{fig:psfSPositiveFocus}
    \end{subfigure}
     \begin{subfigure}[ht]{0.42\textwidth}
        \centering
        \includegraphics[width=\textwidth]{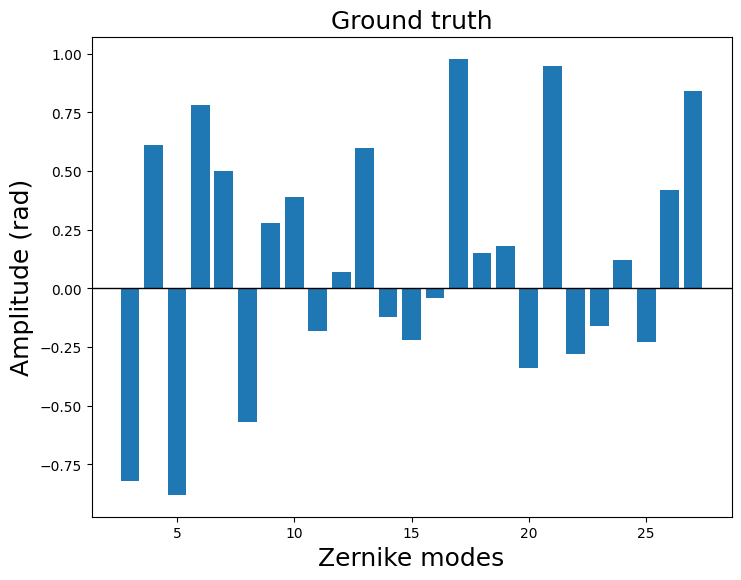}
        \caption{Ground truth Zernike coefficients}
        \label{fig:psfSGT}
    \end{subfigure}
    \begin{subfigure}[ht]{0.32\textwidth}
        \centering
        \includegraphics[width=\textwidth]{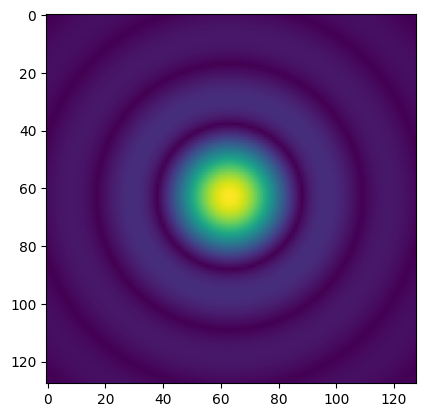}
        \caption{Ground truth sample}
        \label{fig:psfSOri2}
    \end{subfigure}
    \caption{An example of an input simulated \ac{psf} dataset with defocus $Z_4$ as the bias mode providing the phase-diversity. a), b), and c) show simulated images of the PSF at different focal planes when a random set of Zernike modes, as presented in d), are applied. e) shows the ground truth \ac{psf} without any aberration.}
    \label{fig:psfSInput}
\end{figure}

The aberrations are represented by a series of Zernikes polynomials randomly sampled from a uniform distribution within the amplitude range of [-1,1] radians. The coefficients are used to calculate a point spread function, which is used directly to recover a sharp point object or convolved with an extended image to represent an aberrated image. The simulated \ac{psf} images are synthesized using the first 25 Zernike modes $Z_3-Z_{27}$ (ANSI indices), while excluding $Z_0$ (piston), $Z_1$ (tip), and $Z_2$ (tilt), as they can be easily corrected using centroiding algorithms or other registration methods.  A bias mode (e.g. defocus $z_4$) with coefficient 1 rad may be added or subtracted from the Zernike coefficients to generate the phase-diverse images $I_{-1}, I_0, I_{+1}$. A sample simulated \ac{psf} input set is shown in Figure \ref{fig:psfSInput}, consisting of the phase-diverse images alongside the corresponding Zernike coefficients.

\begin{figure}[ht]
    \centering
    \begin{subfigure}[ht]{0.32\textwidth}
        \centering
        \includegraphics[width=\textwidth]{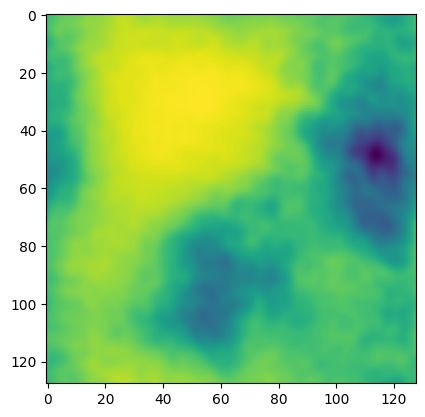}
        \caption{Negative defocus, $I_{-1}$}
        \label{fig:psfexSNegativeDefocus}
    \end{subfigure}
    \hfill
    \begin{subfigure}[ht]{0.32\textwidth}
        \centering
        \includegraphics[width=\textwidth]{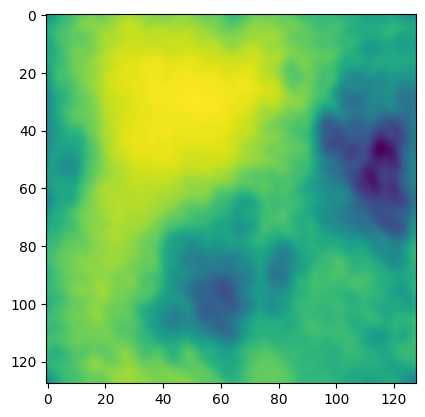}
        \caption{At focus, $I_{0}$}
        \label{fig:psfexSAtFocus}
    \end{subfigure}
    \begin{subfigure}[ht]{0.32\textwidth}
        \centering
        \includegraphics[width=\textwidth]{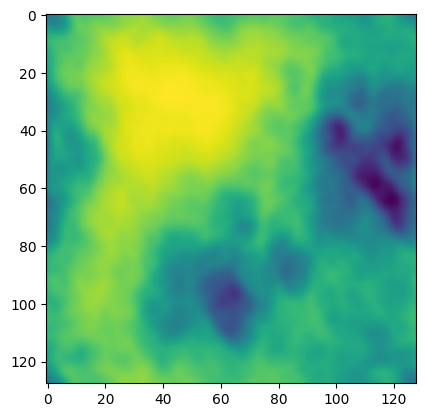}
        \caption{Positive defocus, $I_{+1}$}
        \label{fig:psfexSPositiveFocus}
    \end{subfigure}
     \begin{subfigure}[ht]{0.42\textwidth}
        \centering
        \includegraphics[width=\textwidth]{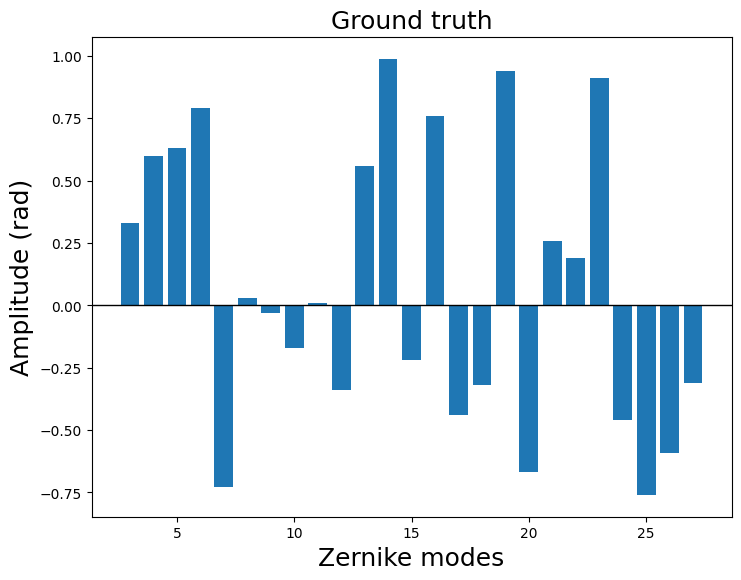}
        \caption{Ground truth Zernike coefficients}
        \label{fig:exSGT}
    \end{subfigure}
    \begin{subfigure}[ht]{0.32\textwidth}
        \centering
        \includegraphics[width=\textwidth]{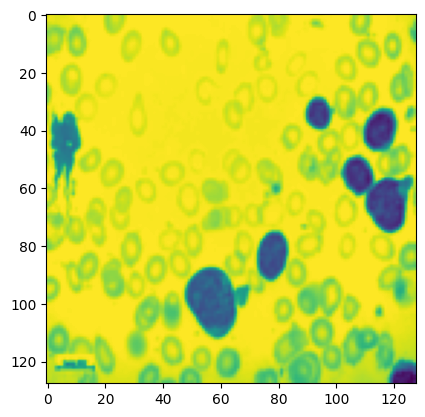}
        \caption{Ground truth sample}
        \label{fig:exSIGT}
    \end{subfigure}
    \caption{An example of an input simulated extended 2D sample dataset with defocus $Z_4$ as the bias mode providing the phase-diversity. a), b), and c) show simulated images of the 2D sample at different focal planes when a random set of Zernike modes, as presented in d), are applied. e) shows the ground truth 2D sample without any aberration.}
    \label{fig:exSInput}
\end{figure}

In the case of the simulated extended 2D sample dataset, we utilised a mixed dataset, including artificially generated textured dataset from class 7 and 10 of the DAGM (Deutsche Arbeitsgemeinschaft für Mustererkennung e.V., the German chapter of the International Association for Pattern Recognition) 2007 competition dataset (n=4600) \cite{wieler2007weakly}, Blood Cells for Acute Lymphoblastic Leukemia dataset (n=3242) \cite{hosseini2023mobile} and Malaria infected human blood smears image set BBBC041v1 available from the Broad Bioimage Benchmark Collection (n=1328) \cite{ljosa2012annotated}. These sample images are then convolved with \ac{psf} kernels at the planes associated with the bias defocus to generate aberrated images for evaluation. An example of the input images from the simulated extended 2D sample dataset is shown in Figure \ref{fig:exSInput}, consisting of the phase-diverse images alongside the corresponding Zernike coefficients.

\subsection{Framework Overview} 
\begin{figure}[ht]
\begin{center}
\includegraphics[width=\textwidth]{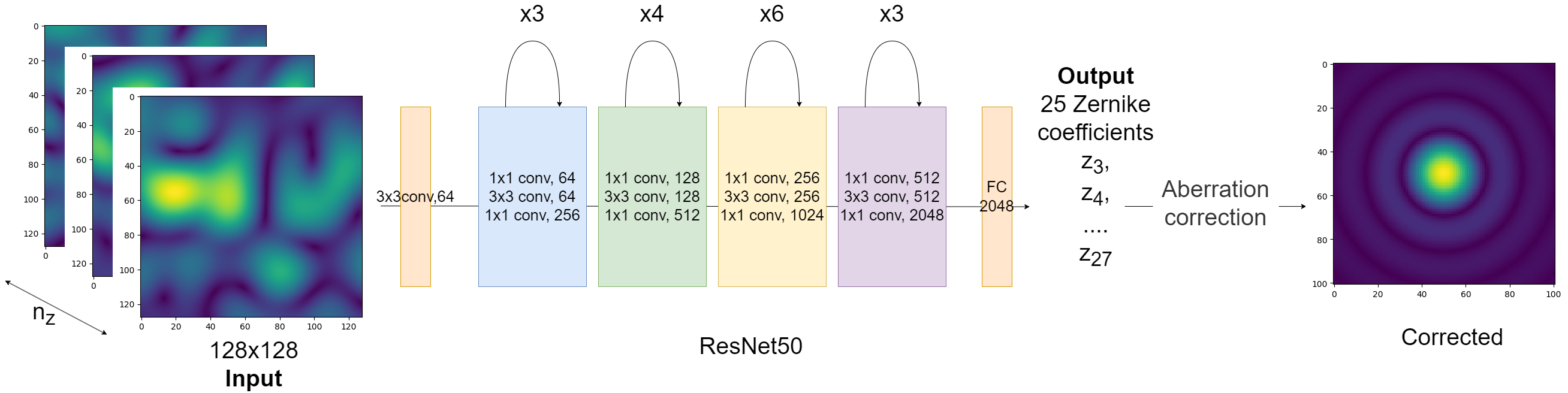}
\caption{Overview of framework: The ResNet50 is trained to predict the Zernike coefficients from  $n_z$ phase-diversity images captured at different planes ($n_z$ not more than three in this study). The predicted Zernike coefficients are used to correct the aberration on the image.}
\label{fig:architecture}
\end{center}
\end{figure}

We utilise a residual network (ResNet50) \cite{he2016deep}, a standard backbone network for various computer vision tasks, to predict the Zernike coefficients from the phase-diverse images. To accommodate the specific requirements of each experiment, we adapted the network to accept $n_z$-channel input, with the value of $n_z$ not exceeding three in this work. Also, the last fully connected layer is modified to output 25 Zernike coefficients. The overall framework is illustrated in Figure \ref{fig:architecture}.

\subsection{Experimental Design} 
For all experiments using the simulated \ac{psf} dataset, we generated 60,000 independent validation samples and 60,000 test samples. For the training set, we employed a dynamic approach in which random Zernike values were used to produce new simulated training samples that changed during each epoch of training, thereby enhancing the diversity of the data. To align with the sizes of the validation and test sets, we consider one epoch to comprise 480,000 training samples. No validation or testing samples were used to train the network. The model was trained across the input images to infer all 25 Zernike coefficients, $Z_3-Z_{27}$, with each being individually limited to an absolute maximum value of 1 radian.

In the experiments involving simulated extended 2D images, we first split the mixed dataset (n=9170) in a ratio of 80:10:10 for the training, validation and test sets. We then duplicate images within each dataset to produce an equivalent 480,000 images for training, 60,000 images each for the validation and test sets, to match the \ac{psf} dataset. Similar to the experiments on the simulated \ac{psf} dataset, the 480,000 \ac{psf} kernel sets were generated on-the-fly during training by convolving across the training images, whereas the images from the validation and test sets were convolved prior to the experiment beginning, and then fixed. All images were transformed into grayscale, and augmented using random flipping, rotation between 0\textdegree and 70\textdegree, translation between 10\% and 30\% of input image resolution, and scaling between 90\% and 110\% of input image resolution to enhance the variability of the images. 

In general, each model is trained for 100 or 250 epochs respectively for the \ac{psf} and extended 2D sample dataset, with a batch size of 64. All images are resized to 128x128 pixels resolution prior to input into the network. The validation set is used to determine the optimal number of training epochs, with the final performance calculated through evaluation on the test set. We used \ac{mse} as the loss function and Adam optimizer with a learning rate of 0.0005, betas (0.9, 0.999) across all experiments. 

To assess the accuracy of the Zernike coefficient predictions, we used the \ac{rmse} as the evaluation metric to quantify the difference between the original and the estimated Zernike coefficients (i.e. the residual wavefront error that would remain in an optical system after correction).

To visually illustrate the extent of correction achievable through Zernike coefficient prediction, we subtracted the predicted Zernike coefficients from the ground truth Zernike coefficients. This resultant subtraction was then applied to the original ground truth image to simulate the correction that can be performed.

Building upon prior works that primarily focused on defocus as the standard bias mode for phase-diverse imaging \cite{ma2019numerical,guo2019improved,andersen2019neural,andersen2020image,wu2020sub,krishnan2020optical,saha2020practical,ma2023turbulence,shohani2023using}, as well as the work by Hu et al. \cite{hu2023universal} that proposes the astigmatism mode as the most effective bias mode, we explored the use of different bias modes. Given the inherent ambiguity associated with single-channel inputs, we examined the minimum number of phase-diverse images required for better performance, and investigated the different combinations of amplitudes on the aberrated \ac{psf} images. Inspired by iterative optimisation algorithms used in existing adaptive optics systems, we also simulated a short iterative process to determine whether progressively correcting aberrations within the \ac{psf} images enhances performance. To evaluate the network's potential real-world applicability, we also evaluate the approach on extended 2D sources. The experiments were as follows:

\noindent\textbf{\emph{E1}}: Using the simulated \ac{psf} dataset, we vary the bias mode employed to generate our standard 3-channel phase-diverse images, $I_{-1}, I_0, I_{+1}$. The considered bias modes include the oblique astigmatism $z_3$, defocus $z_4$, primary spherical $z_12$, and vertical secondary trefoil $z_{16}$ modes. For comparison purposes, we also present a baseline model that utilises a single-channel input without any bias modes.

\noindent\textbf{\emph{E2}}: Using the simulated \ac{psf} dataset, we investigate whether a 2-channel input is sufficient to predict the Zernike coefficients and explore combinations of defocus mode's amplitude (negative, positive or at focus) for the 2-channel input. Defocus is used here as the best performing bias mode from \emph{E1}. To facilitate comparison, we evaluate the performance of the model with the 2-channel input against the model with the 3-channel input from our earlier experiment.

\noindent\textbf{\emph{E3}}: We simulate an iterative correction process by repeatedly removing the predicted Zernikes from aberrated images. First, the network predicts the Zernike coefficients present in the input \ac{psf}. These predicted Zernike coefficients are then subtracted from the ground truth Zernike coefficients to determine the residual aberration. Finally, the calculated residual aberration is then re-applied to the ground truth \ac{psf}, simulating a corrected image. This process is iterated a small number of times to establish whether iterative correction in this setting is effective.

\noindent\textbf{\emph{E4}}: Utilising the optimal parameters obtained from \emph{E1} and \emph{E2}, we evaluate the model on the simulated extended 2D sample dataset.

\section{Results and Discussion}
\subsection{Simulated \ac{psf} dataset}

\textbf{Bias Mode:} Based on table \ref{table:biasRes}, the network using the defocus mode demonstrates the best performance, exhibiting an improvement of approximately 80\% compared to the baseline single-channel input. This finding aligns with prior research\cite{siddik2023deep,saha2020practical}, which highlights the challenges associated with single-image ambiguity for accurate Zernike coefficient prediction. On the other hand, the network employing the oblique astigmatism mode underperforms compared to defocus, even though previous study\cite{hu2023universal} has reported oblique astigmatism to be more effective when predicting for smaller range of Zernike coefficients (N=5-9) from pre-processed pseudo-\ac{psf} derived from phase-diverse images. Our interpretation is that when directly predicting on large range of Zernike coefficients (N=25), the defocus mode serves as a more appropriate bias mode compared to the other modes. To explore the potential improvement offered by higher-order bias modes, we evaluated the network using primary spherical $z_{12}$ and vertical secondary trefoil mode, $z_{16}$. However, our results did not support this hypothesis, suggesting that higher-order bias modes may not yield notable improvements.
    
\begin{table}[ht]
\centering
\caption{Comparison of using the different bias mode for the 3-channel phase-diverse input on the simulated \ac{psf} dataset.}
\label{table:biasRes}
\begin{tabular}{@{ }l|c@{ }} 
\toprule
\textbf{Bias mode}                   & \textbf{RMSE}  \\ 
\midrule
Oblique Astigmatism, $z_3$        &     0.1468           \\ 

Defocus, $z_4$                    &    0.1020            \\ 

Primary spherical, $z_{12}$ &      0.1797         \\ 

Vertical secondary trefoil, $z_{16}$ &      0.4815         \\ 

None/single-channel (baseline)                      &    0.5019           \\
\bottomrule
\end{tabular}
\end{table}

\noindent{\textbf{Channel Inputs:}} 2-channel input (different combinations of amplitudes) or a 3-channel input
Results on Table \ref{table:channelRes} show that the 3-channel phase-diverse input remain superior over other 2-channel input with different combinations of amplitudes. The 2-channel input with the combination of negative defocus and at focus plane [-1,0] as well as the combination of at focus and positive defocus planes [0,1] demonstrate performances that are slightly inferior to the 3-channel input. This observation suggests that the inclusion of all three channels in the input enables the network to discern the inter-channel distinctions necessary for determining Zernike coefficient values. Conversely, the combination of negative and positive defocus planes [-1,1] yields the worst results, suggesting that the information contained within the at focus plane is the key to the Zernike prediction.

\begin{table}[ht]
\centering
\caption{Comparison of using the different amplitudes of the defocus mode within the 3-channel input on the simulated \ac{psf} dataset.}
\label{table:magnitudeRes}
\begin{tabular}{@{ }l|c@{ }}
\toprule
\textbf{Amplitudes of the defocus mode, $z_4$ (channels) } & \textbf{RMSE}  \\ 
\midrule
{[}-0.5, 0, 0.5]                            & 0.1031         \\ 

{[}-1, 0, 1]                                 & 0.1020         \\ 

{[}-1.5, 0, 1.5]                            & 0.1439         \\
\bottomrule
\end{tabular}
\end{table}

\begin{table}[ht]
\centering
\caption{Comparison of using a 2-channel input with different combinations of amplitudes and a 3-channel input on the simulated \ac{psf} dataset.}
\label{table:channelRes}
\begin{tabular}{@{ }c|l|c@{ }}
\toprule
\multicolumn{2}{c|}{Defocus mode, $z_4$ (channels)} & \textbf{RMSE} \\
\midrule
\multirow{3}{*}{\textbf{2-channel}} & {[}-1,0] & 0.1210 \\
& {[}0,1] &  0.1248 \\
& {[}-1,1] &  0.1786 \\
\midrule
\textbf{3-channel} & {[}-1,0,1] &  0.1020 \\
% Defocus mode, $z_4$ (channels) & {[}-1,0]            & {[}0,1] & {[}-1,1] & {[}-1,0,1]         \\
% RMSE                          & 0.1210              & 0.1248  & 0.1786   & 0.1020             \\
\bottomrule
\end{tabular}
\end{table}

We conclude that a phase-diverse input, captured above, below and at the focal plane with an amplitude of 1 rad yield the best results for the simulated \ac{psf} dataset. Figure \ref{fig:psfResult} presents qualitative results obtained by using the network with the optimal parameters. Here we simulate a single-shot adaptive optics solution by rendering the corrected \ac{psf} having subtracted the predicted Zernike from the original randomly applied aberration. The figures show that a large amount of the aberration has been successfully removed, and that the final \ac{psf} is closer to the true ground truth. Some small artefacts remain, primarily on the lobes of the \ac{psf} due to some small residual Zernike values that have not been completely corrected. In Appendix A, we provide additional numerical results, specifically \ac{psnr} and \ac{ssim} values showing a good quantitative similarity between the corrected \ac{psf} and the ground truth, as well as illustrations of the numerical similarity between the random ground truth and predicted Zernike coefficients.

\newlength{\subfigwidth}
\setlength{\subfigwidth}{45mm}
\begin{figure}[ht]
\captionsetup[subfigure]{labelformat=empty}
\centering
\begin{tabular}{p{\subfigwidth} p{\subfigwidth} p{\subfigwidth}}

\begin{subfigure}[b]{\subfigwidth}
    \caption{Ground truth} 
    \includegraphics[width=\subfigwidth]{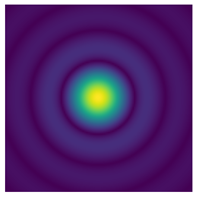}
\end{subfigure} &
\begin{subfigure}[b]{\subfigwidth}
    \caption{Aberrated}
    \includegraphics[width=\subfigwidth]{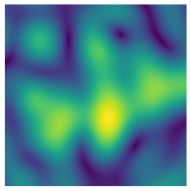}
\end{subfigure} &
\begin{subfigure}[b]{\subfigwidth}
    \caption{Corrected \ac{psf}}
    \includegraphics[width=\subfigwidth]{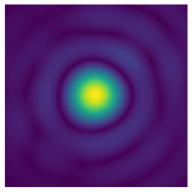}
\end{subfigure} \\

\includegraphics[width=\subfigwidth]{images/rg1.png} &
\includegraphics[width=\subfigwidth]{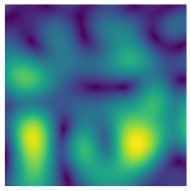} &
\includegraphics[width=\subfigwidth]{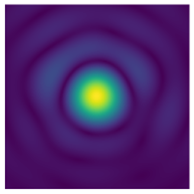} \\

\includegraphics[width=\subfigwidth]{images/rg1.png} &
\includegraphics[width=\subfigwidth]{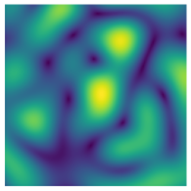} &
\includegraphics[width=\subfigwidth]{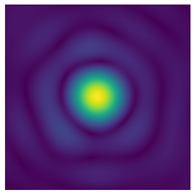} \\

\end{tabular}
\caption{Subset of results obtained on the simulated \ac{psf} dataset. From the left to right columns are the ground truth images, the aberrated images and the corrected images after Zernike prediction. In this simulated experiment, the corrected image is calculated from the residual aberration (Ground truth Zernike values minus Predicted Zernikes) re-applied to the ground truth image.}
\label{fig:psfResult}
\end{figure}

\noindent{\textbf{Iterative correction:}} We show the qualitative output of an iterative aberration correction approach in Figure \ref{fig:iterativeRes}. We find that the iterative correction process requires only two or three iterations to further reduce the residual aberration present. This improvement suggests the network can predict and correct even very small aberrations within the image. This iterative approach offers the potential to refine image correction progressively, potentially mitigating errors even if the initial prediction is inaccurate, however we also note that a single step remains effective.

\setlength{\subfigwidth}{30mm}
\begin{figure}[ht]
\captionsetup[subfigure]{labelformat=empty}
\centering
\begin{tabular}{p{\subfigwidth} p{\subfigwidth} p{\subfigwidth} p{\subfigwidth}p{\subfigwidth}}

 & &\multicolumn{3}{c}{Corrected PSF} \\

\begin{subfigure}[b]{\subfigwidth}
    \caption{Ground truth} 
    \includegraphics[width=\subfigwidth]{images/rg1.png}
\end{subfigure} &
\begin{subfigure}[b]{\subfigwidth}
    \caption{Aberrated}
    \includegraphics[width=\subfigwidth]{images/ra2.png}
\end{subfigure} &
\begin{subfigure}[b]{\subfigwidth}
    \caption{1$^{st}$ iteration}
    \includegraphics[width=\subfigwidth]{images/rp2.png}
\end{subfigure} &
\begin{subfigure}[b]{\subfigwidth}
    \caption{2$^{nd}$ iteration}
    \includegraphics[width=\subfigwidth]{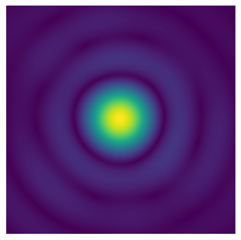}
\end{subfigure} &
\begin{subfigure}[b]{\subfigwidth}
    \caption{3$^{rd}$ iteration}
    \includegraphics[width=\subfigwidth]{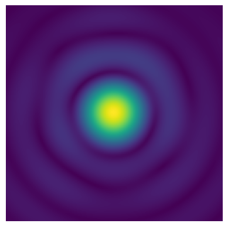}
\end{subfigure}
\\

\includegraphics[width=\subfigwidth]{images/rg1.png} &
\includegraphics[width=\subfigwidth]{images/ra3.png} &
\includegraphics[width=\subfigwidth]{images/rp3.png} &
\includegraphics[width=\subfigwidth]{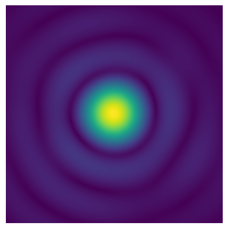} &
\includegraphics[width=\subfigwidth]{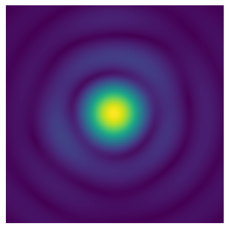} \\

\end{tabular}
\caption{Iterative Zernike coefficient prediction on the simulated \ac{psf} dataset. From the left to right columns are the ground truth images, the aberrated images and the corrected images after Zernike prediction at 1$^{st}$, 2$^{nd}$ and 3$^{rd}$ iterations. At the 1$^{st}$ iteration, the corrected image is calculated from the residual aberration (Ground truth Zernike values minus Predicted Zernikes) re-applied to the ground truth image. At each subsequent iteration, the network predicts the aberration remaining in the previous corrected image and the calculated residual aberration is again applied on the ground truth image.}

\label{fig:iterativeRes}
\end{figure}

\subsection{Simulated extended 2D sample dataset}
For the extended 2D sample dataset, we perform a similar experiment to that of the simulated \ac{psf}, except we only choose the best performing bias mode, defocus, across all three channels. Table~\ref{table:simExtendedRes}) shows good performance of the Zernike prediction network on this task. The \ac{rmse} increases slightly for the simulated extended 2D sample compared to the \ac{psf} dataset, likely representing the substantial increase in challenge of this dataset over a clean simulated dataset. The performance nevertheless remains good when correcting most aberrations across the images we have tested. Figure \ref{fig:extendedResult} showcases a subset of qualitative results obtained on the simulated extended 2D sample dataset. As with our experiments on \ac{psf} images, we simulate correction of the image by applying a new set of corrected Zernike to the original ground truth image. Consistent with the performance observed on the \ac{psf} dataset, the visual corrections to the image are able to restore much of the lost quality observed under substantial random aberrations. This is further supported by the high similarity values and accurate Zernike coefficient predictions detailed in Appendix B.

\begin{table}[ht]
\centering
\caption{Evaluation of the ResNet50 model on the simulated extended 2D sample dataset using a 3-channel phase-diverse input captured at the defocus plane.}
\label{table:simExtendedRes}
\begin{tabular}{@{ }l|c@{ }}
\toprule
\textbf{Amplitudes of the defocus mode, $z_4$ (channels) } & \textbf{RMSE}  \\ 
\midrule
{[}-1, 0, 1]                            & 0.1479         \\
\bottomrule
\end{tabular}
\end{table}

\setlength{\subfigwidth}{45mm}
\begin{figure}[ht]

\captionsetup[subfigure]{labelformat=empty}
\centering
\begin{tabular}{p{\subfigwidth} p{\subfigwidth} p{\subfigwidth}}

\begin{subfigure}[b]{\subfigwidth}
    \caption{Ground truth} 
    \includegraphics[width=\subfigwidth]{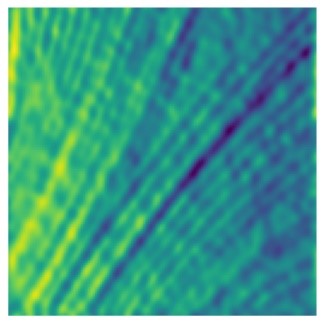}
\end{subfigure} &
\begin{subfigure}[b]{\subfigwidth}
    \caption{Aberrated}
    \includegraphics[width=\subfigwidth]{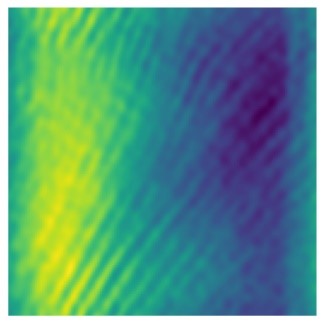}
\end{subfigure} &
\begin{subfigure}[b]{\subfigwidth}
    \caption{Corrected Image}
    \includegraphics[width=\subfigwidth]{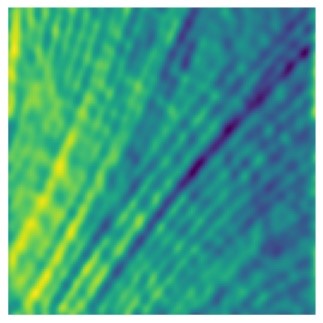}
\end{subfigure} \\

\includegraphics[width=\subfigwidth]{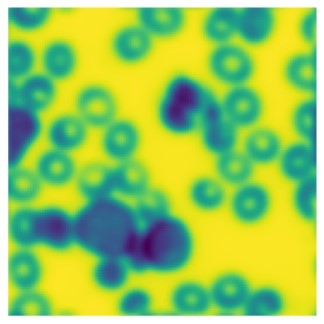} &
\includegraphics[width=\subfigwidth]{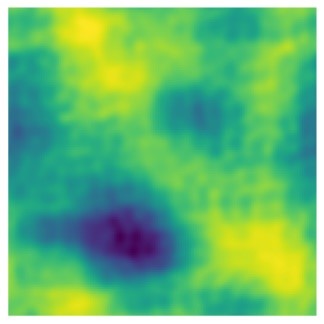} &
\includegraphics[width=\subfigwidth]{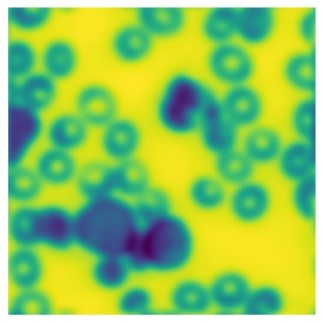} \\

\includegraphics[width=\subfigwidth]{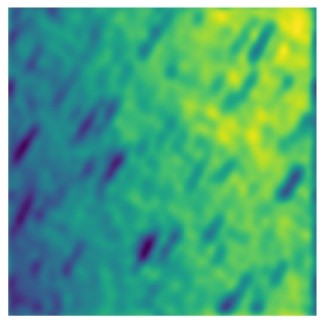} &
\includegraphics[width=\subfigwidth]{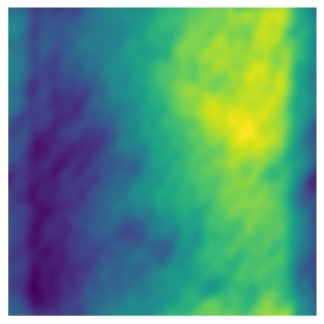} &
\includegraphics[width=\subfigwidth]{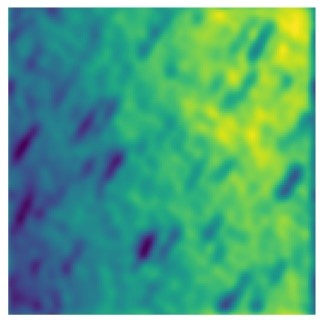} \\

\includegraphics[width=\subfigwidth]{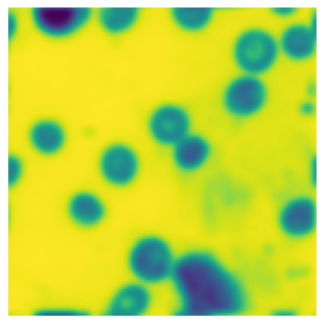} &
\includegraphics[width=\subfigwidth]{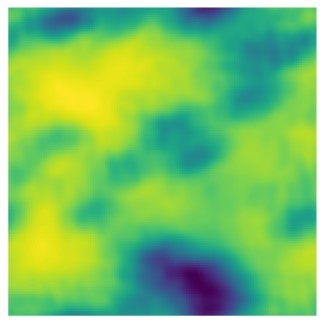} &
\includegraphics[width=\subfigwidth]{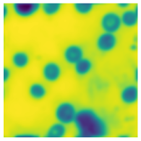} 
\end{tabular}
\caption{Qualitative results obtained on the simulated extended 2D sample dataset. From the left to right columns are the ground truth images, the aberrated images and the corrected images after Zernike prediction and removal. As with the \ac{psf} simulated results, the corrected image is calculated from the residual aberration (Ground truth Zernike values minus Predicted Zernikes) re-applied to the ground truth image.}
\label{fig:extendedResult}
\end{figure}

\section{Conclusion}
\label{sec:conclusion}
This study has demonstrated the application of a ResNet50 model for direct prediction of optical aberrations as Zernike coefficients in both simulated \ac{psf} and extended 2D sample datasets. Utilising predictions from the model, this approach could enable rapid aberration correction within a microscope system. The approach avoids an iterative solution that requires many optimisation steps, possibly with additional image capture, and requires no prior knowledge of the present aberrations beyond simply an aberrated image. The approach requires only two or three images captured at different focal planes (above, below, and in focus), theoretically enabling rapid image capture and correction when deployed in real systems. This could potentially reduce both acquisition time and specimen exposure. The evaluation, encompassing 25 Zernike coefficients, suggests that the approach remains effective for large aberration across a broad range of Zernike polynomials. The approach is effective with only a single iteration of prediction from the deep network, we also show that additional iterations can further reduce aberration with minimal overhead. Future work will explore the potential of this approach to real-world scenarios through evaluation on experimental microscopy datasets.

\clearpage
\bibliography{aberration}

\end{document}